\documentclass[aip,rsi,amsmath,amssymb,reprint]{revtex4-1}
\usepackage{diagbox}

\usepackage{graphicx}% Include figure files
\usepackage{dcolumn}% Align table columns on decimal point
\usepackage{bm}% bold math

\usepackage[utf8]{inputenc}
\usepackage[T1]{fontenc}
\usepackage{mathptmx}
\usepackage{etoolbox}

\usepackage{amssymb,amsmath,placeins,epsfig,array}
\usepackage[usenames,dvipsnames]{color}
\usepackage[normalem]{ulem}
\usepackage[breaklinks,colorlinks,urlcolor=blue,citecolor=blue,linkcolor=blue]{hyperref}
\usepackage{lineno}
\usepackage{mathptmx}
\usepackage{scalerel}
\usepackage{tikz}
\usetikzlibrary{svg.path}

\definecolor{orcidlogocol}{HTML}{A6CE39}
\tikzset{
  orcidlogo/.pic={
    \fill[orcidlogocol] svg{M256,128c0,70.7-57.3,128-128,128C57.3,256,0,198.7,0,128C0,57.3,57.3,0,128,0C198.7,0,256,57.3,256,128z};
    \fill[white] svg{M86.3,186.2H70.9V79.1h15.4v48.4V186.2z}
                 svg{M108.9,79.1h41.6c39.6,0,57,28.3,57,53.6c0,27.5-21.5,53.6-56.8,53.6h-41.8V79.1z M124.3,172.4h24.5c34.9,0,42.9-26.5,42.9-39.7c0-21.5-13.7-39.7-43.7-39.7h-23.7V172.4z}
                 svg{M88.7,56.8c0,5.5-4.5,10.1-10.1,10.1c-5.6,0-10.1-4.6-10.1-10.1c0-5.6,4.5-10.1,10.1-10.1C84.2,46.7,88.7,51.3,88.7,56.8z};
  }
}

\newcommand\orcidicon[1]{\href{https://orcid.org/#1}{\mbox{\scalerel*{
\begin{tikzpicture}[yscale=-1,transform shape]
\pic{orcidlogo};
\end{tikzpicture}
}{|}}}}

\usepackage{graphicx}% Include figure files
\usepackage{dcolumn}% Align table columns on decimal point
\usepackage{bm}% bold math
\usepackage{subfigure}
\usepackage{epstopdf}
\usepackage{color}
\usepackage[utf8]{inputenc}

\makeatletter
\def\@email#1#2{%
 \endgroup
 \patchcmd{\titleblock@produce}
  {\frontmatter@RRAPformat}
  {\frontmatter@RRAPformat{\produce@RRAP{*#1\href{mailto:#2}{#2}}}\frontmatter@RRAPformat}
  {}{}
}%
\makeatother

\begin{document}
\preprint{AIP/123-QED}
\title{\large A rotation-based Hall-probe magnetic compass}

\author {B.~Wojtsekhowski \orcidicon{0000-0002-2160-9814}} 
\email[Contact person, ]{bogdanw@jlab.org} 
\affiliation{Thomas Jefferson National Accelerator Facility,  Newport  News, Virginia 23606, USA}
\author{C.~Cuevas}
\affiliation{Thomas Jefferson National Accelerator Facility,  Newport  News, Virginia 23606, USA}
\author{R.~Dawson}
\affiliation{Thomas Jefferson National Accelerator Facility, Newport  News, Virginia 23606, USA}
\author{W.~Henry}
\affiliation{Thomas Jefferson National Accelerator Facility, Newport  News, Virginia 23606, USA}
\author{N.~Liyanage}
\affiliation{University of Virginia, Charlottesville, Virginia 22904, USA}
\author{G.~Ron}
\affiliation{\mbox{The Hebrew University of Jerusalem{,} Jerusalem 9190401, Israel}}
\author{J.~Wilson}
\affiliation{Thomas Jefferson National Accelerator Facility, Newport  News, Virginia 23606, USA}
\begin{abstract}
In many physics and space experiments the direction of a magnetic field must be known to one milliradian or better.
We have developed a new type of magnetic compass, based on rotation, in which a spinning Hall probe (HP) produces an alternating signal proportional to the component of the magnetic field transverse to the axis of rotation.
Aligning the rotation axis so that this signal vanishes gives the direction of the field.
The device requires no calibration and is insensitive to electronics drift.
A spinning-HP compass was built and used in a recent experiment at Jefferson Lab, where it determined the direction of the 25~G holding field of a polarized $^3$He target to about 1~mrad; the direction of the rotation axis was transferred to the laboratory frame with a laser and a flat mirror attached to the rotor.
A second implementation of the rotation approach, a spinning field concentrator (SFC) with a stationary Hall probe, was also built.
With 10 seconds of averaging the SFC compass resolves a transverse field of 18~$\mu$G, corresponding to a directional resolution of 70~$\mu$rad in a field of 0.25 G, close to the Earth's field.
\end{abstract}

\date{\today}

\maketitle
%    \linenumbers
\section{Introduction}
\label{sec:introduction}
The magnetic compass has been used for navigation at sea and on land for about a thousand years~\cite{CMC}.
The classical compass is a magnetized needle acted on by the torque $\vec T = \vec M \times \vec B$ between its magnetic moment $\vec M$ and the field $\vec B$.
The torque vanishes when $\vec M$ is parallel to $\vec B$, thus the needle points along the field direction.
For best precision the pivot must have as little friction as possible, but even so an accuracy much better than one degree is difficult to reach.
A further systematic error arises because the magnetic moment of the needle is in general not exactly parallel to its geometrical axis.

Almost 130 years ago A.~Righi, after discovering a huge Hall effect in Bismuth, built the first Hall-effect-based Earth magnetic compass, see Fig.~19 in Ref.~\cite{Campbell}.
With the advent of solid-state devices the Hall-probe compass was studied further~\cite{Ross_1957}.
Recent progress in field-direction instrumentation has come mainly from three-axis Hall sensors, see e.g. Ref.~\cite{3D}.
Such sensors reach an angular accuracy of 130~$\mu$rad, but their operation is complicated by electronics drift and by the need for regular calibration, since the orientation of the sensing plane is not known a priori and the probe signals drift.

Rotating-coil detectors are widely used to map the fields of accelerator magnets and spectrometers~\cite{Workshop-2014}, but not for precise determination of the field direction.

\section{Motivation}
The present device was motivated by experiments at Jefferson Lab~\cite{Barabanov, wojtsekhowski_flavor_2020},
in which a polarized $^3$He target is held in a 25~G magnetic field~\cite{He-3}.
The neutron electric form factor at high momentum transfer is measured with a double-polarization method~\cite{Woloshyn}, which requires the direction of the neutron polarization to be known with high accuracy~\cite{GEn-1, GEn-2}.
The uncertainty of the polarization direction propagates into the extracted form factor due to a large ratio of the magnetic to the electric form factors, so the direction of the holding field must be known to about a few mrad.

\section{Geometrical concept of the magnetic compass}
\label{sec:concept}
\begin{figure}[ht]
	\centering
	\includegraphics[trim = 0 0 0 0, width=.95\columnwidth, angle =0]{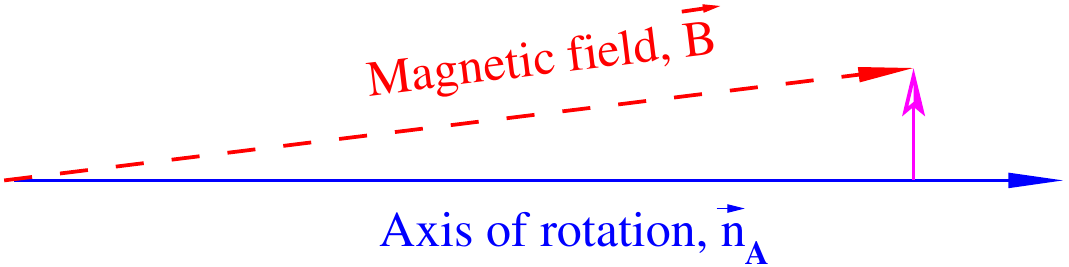}
	\caption{The magnetic field $\vec B$ (red) and its component transverse to the rotation axis (magenta). The unit vector $\vec n_{_A}$ (blue) is along the rotation axis.}
\label{fig:vectors}
\end{figure}

We propose to detect the component of the magnetic field perpendicular to the axis of rotation with a rotating Hall probe, which converts that component into an alternating signal, as shown in Fig.~\ref{fig:vectors}; see also the patent description~\cite{Patent-1}.

The plane of the probe is approximately parallel to the axis of rotation, so the amplitude of the alternating signal is proportional to the transverse field component $|\vec B \times \vec n_{_A}| = B\sin\alpha$, where $\alpha$ is the angle between $\vec B$ and the axis.
Let $S$ be the probe sensitivity, $\beta$ the angle by which the probe normal deviates from the plane perpendicular to the rotation axis, $B_\parallel$ and $B_\perp$ the components of $\vec B$ along and transverse to $\vec n_{_A}$, $\varphi_{_B}$ the azimuth of the transverse component, and $V_0(t)$ a slowly varying offset (probe offset, drift and pickup).
The probe signal is then
\begin{equation}
V(t) = S\left[B_\parallel\sin\beta + B_\perp\cos\beta\,\cos(\Omega t-\varphi_{_B})\right] + V_0(t),
\label{eq:signal}
\end{equation}
where $\Omega$ is the angular frequency of the rotation.
The amplitude of the component at $\Omega$ is $V_\Omega = S B_\perp\cos\beta$, which vanishes only when $B_\perp = 0$, whatever the values of $S$, $\beta$ and $V_0$.
The null therefore defines the direction of the field without any calibration of the probe and without sensitivity to its drift or to its exact orientation on the rotor.

\begin{figure}[ht]
	\centering
	\includegraphics[trim = 0 60 0 20, width=1. \columnwidth, angle =0 ] {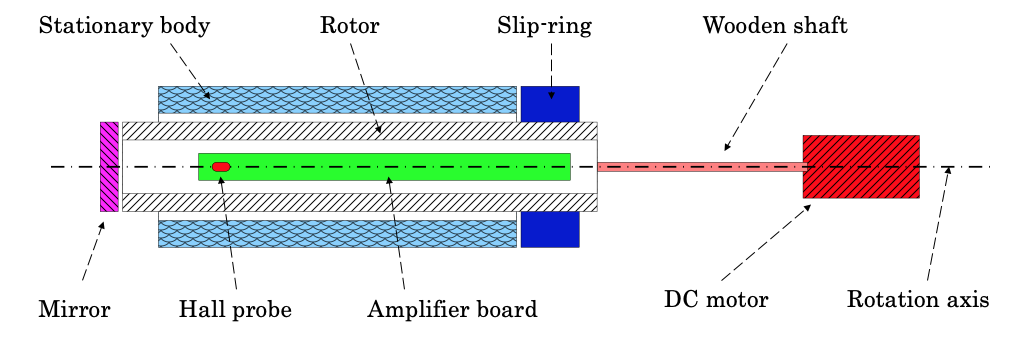}
	\caption{Layout of the spinning-Hall-probe compass (not to scale).}
\label{fig:concept}
\end{figure}

Phase-sensitive detection of this signal, synchronized to the rotation (in the first application by triggered averaging on an oscilloscope), makes the device very accurate: the signal accumulates coherently while the relative contribution of the electronic noise falls with the number of turns~\cite{SPIN23}.
The concept was implemented with the Hall probe located close to the rotor axis; the configuration is shown in
Fig.~\ref{fig:concept}.

\begin{figure}[ht]
	\centering
	\includegraphics[trim = 0 30 0 0, width=1.\columnwidth, angle =0 ] {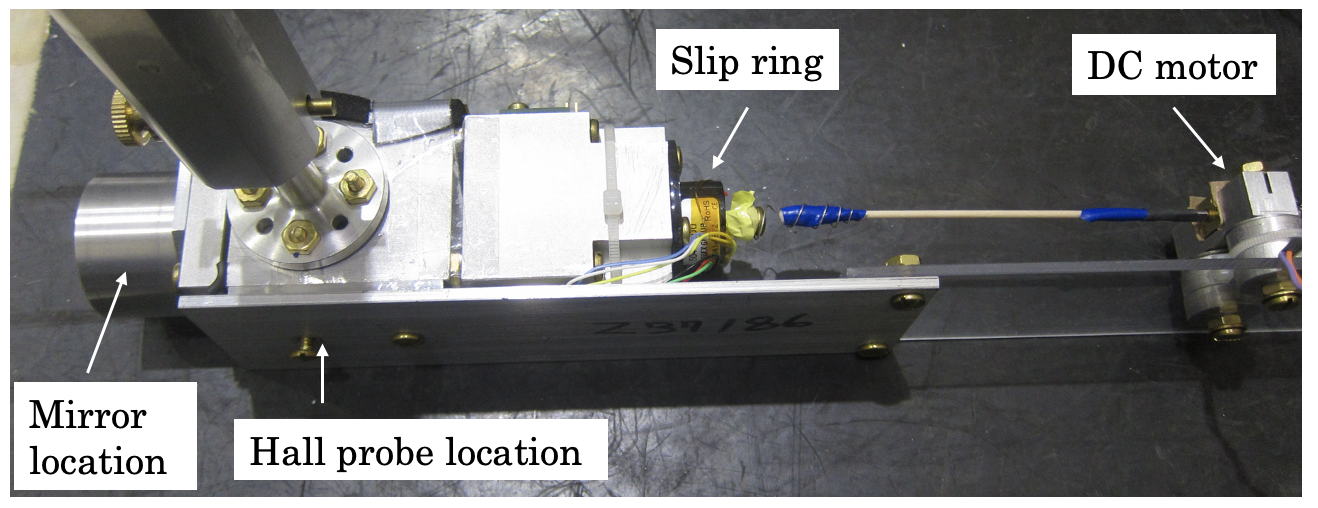}
	\caption{The spinning-Hall-probe compass. The rotor with the Hall probe and the mirror is at left; the DC motor is about 20~cm away along the rotation axis.}
\label{fig:compass}
\end{figure}

A photograph of the device is shown in Fig.~\ref{fig:compass}.

\section{Components of the spinning Hall probe compass}

\begin{figure}[ht]
	\centering
	\includegraphics[trim = 0 30 0 10, width=1. \columnwidth, angle =0 ] {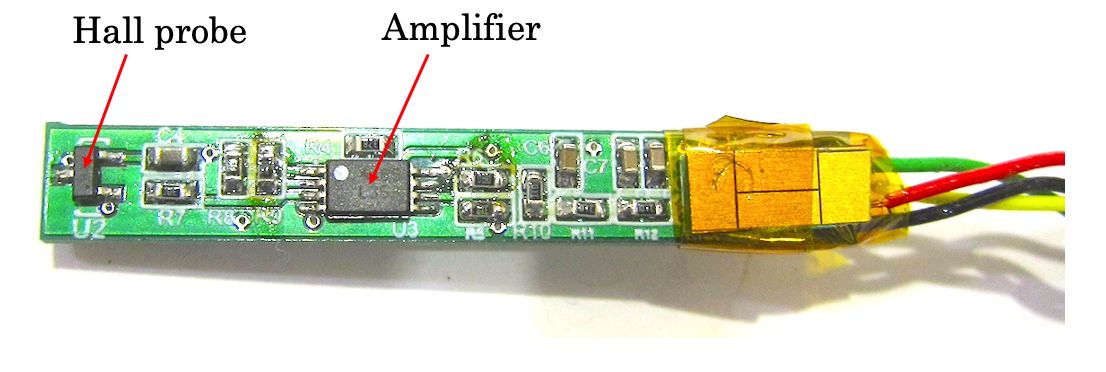}
	\caption{Printed-circuit board carrying the Hall probe and the two-stage front-end amplifier.}
\label{fig:probe}
\end{figure}

Hall sensors are now available at very low cost with sensitivities of up to 9~mV/G (with on-chip amplification)~\cite{Probe}.
Slow drift of the sensor output does not affect the proposed device, since instead of calibrating the HP, we continuously reverse its orientation by rotation.
The probe is mounted on a printed-circuit board (Fig.~\ref{fig:probe}) close to the rotor axis, with its plane approximately parallel to the axis of rotation.

The output signal connection and power to the probe and electronics are provided via a slip-ring~\cite{SlipRing}.
A two-stage amplifier on the board (Fig.~\ref{fig:probe}) raises the signal by a factor of 6, reduces the sensitivity to fluctuations of the slip-ring contact resistance, matches the probe output to the high input impedance of the readout (here an oscilloscope), and restricts the bandwidth to the range of interest (10~Hz).

The rotor is separated from the stationary body of the compass by a pair of precision ceramic bearings.
The slip-ring is also attached to the stationary body, see Fig.~\ref{fig:compass}.

A photodetector on the stationary body provides the reference pulses that mark the rotation phase.

The rotor, which houses the Hall probe, can be driven by a piezoelectric motor~\cite{P-E_motor}, an air turbine, or a DC motor~\cite{DC_motor}.
In the present device a DC motor was used, placed about 20~cm from the Hall probe along the rotation axis (Fig.~\ref{fig:compass}) so that its stray field at the probe is negligible.
The rotation frequency used in experiments was between 5 and 10~Hz.

The optical system for determining the direction of the rotation axis consists of a 405~nm laser pointer, a flat mirror attached to the rotor, and two screens, one between the laser and the compass and one behind the compass.
The front screen is semi-transparent and carries a fine grid on which the positions of the light spots are read.

\begin{figure}[ht]
	\centering
	\includegraphics[trim = 0 0 0 0, width=.85\columnwidth, angle =0 ] {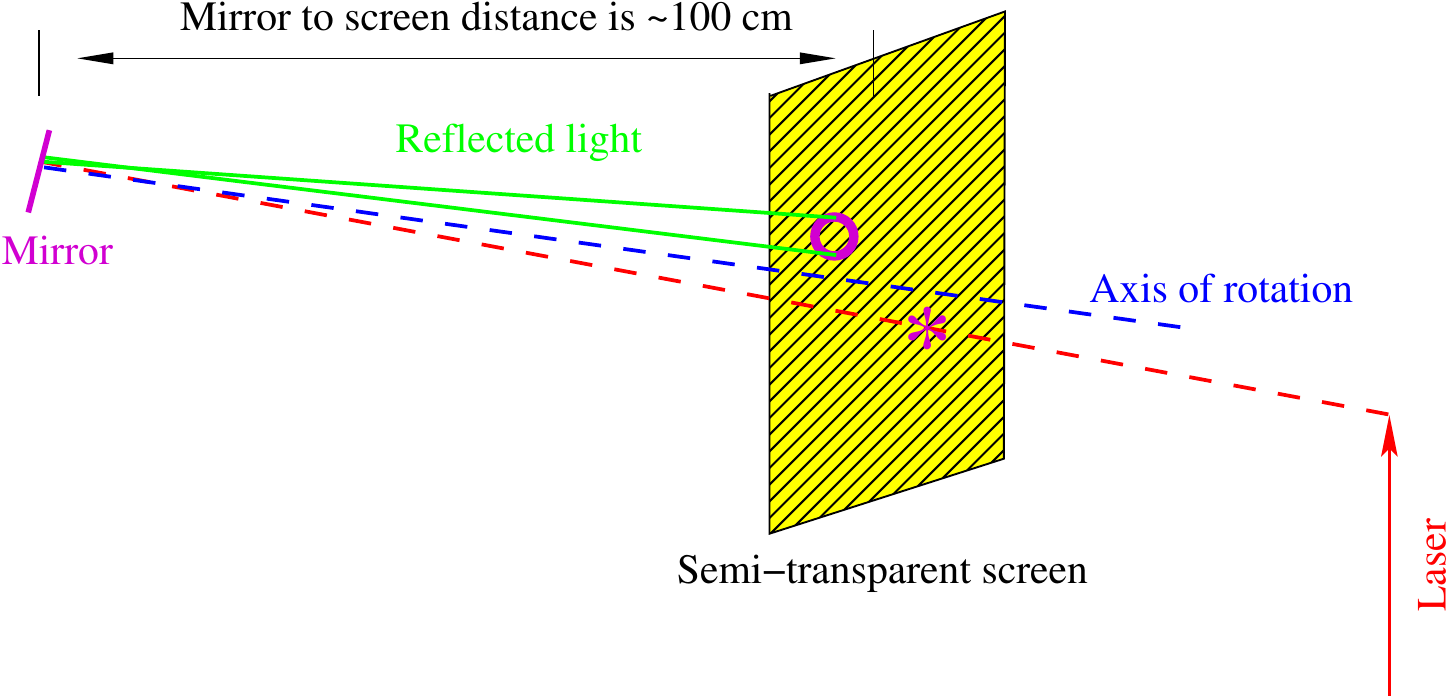}
    \vskip 0.125 in
		\includegraphics[trim = 0 20 0 20, width=.95\columnwidth, angle =0 ] {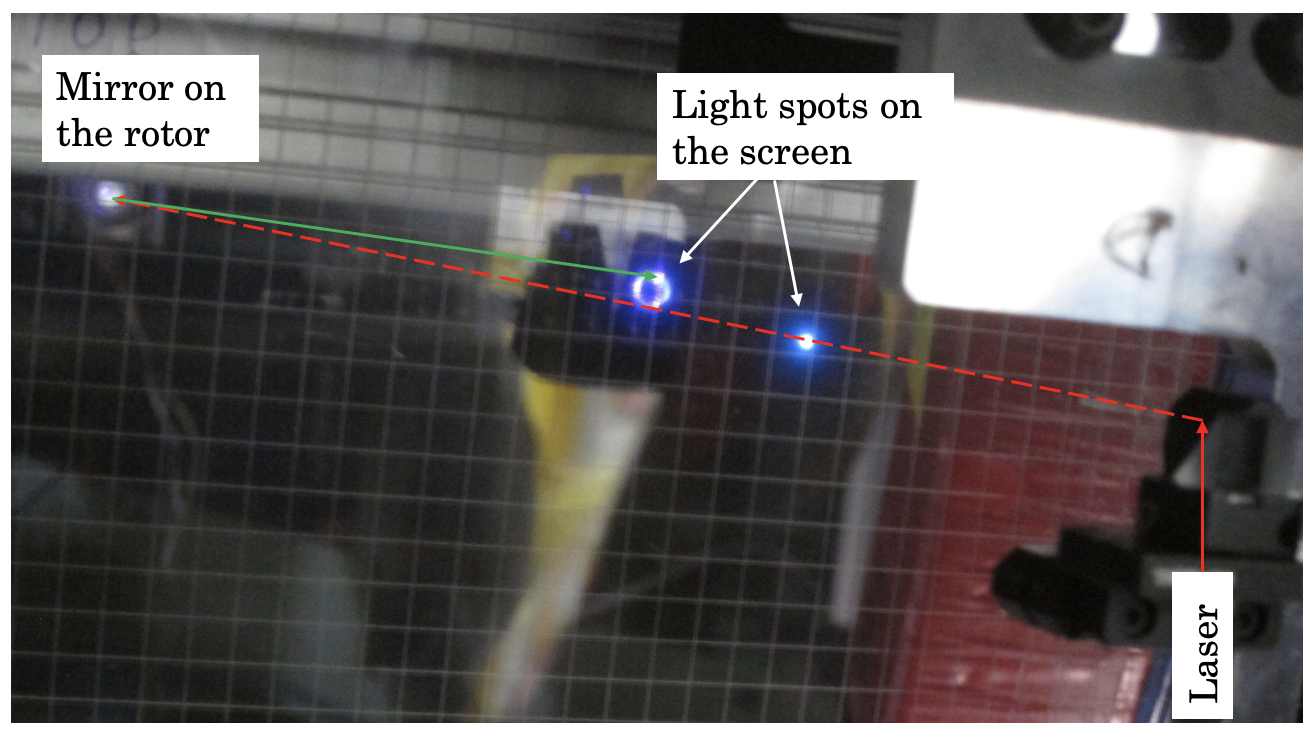}
	\caption{Optical determination of the rotation-axis direction.
	Top: layout. Bottom: photograph of the semi-transparent screen.
	The position of the incident laser spot and the center of the ring of light reflected from the mirror are read from the grid on the screen, whose position was surveyed.
	The reflected light traces a ring because the mirror is slightly non-orthogonal to the rotation axis.
	The red and green lines are added for comparison with the layout above.}
\label{fig:optics}
\end{figure}

The layout and a photograph of the optical system are shown in Fig.~\ref{fig:optics}.
The midpoint between the incident laser spot and the center of the reflected ring gives the direction of the rotation axis.
The second screen allows us to find the direction of the laser light (when the compass is shifted).
The surveyed positions of the screens and the mirror, together with the three spot positions, then fully determine the direction of the rotation axis.

\section{Measurement of the field direction}
	
\begin{figure}[ht]
	\centering
	\includegraphics[trim = 0 0 0 0, width=0.85\columnwidth, angle =0 ] {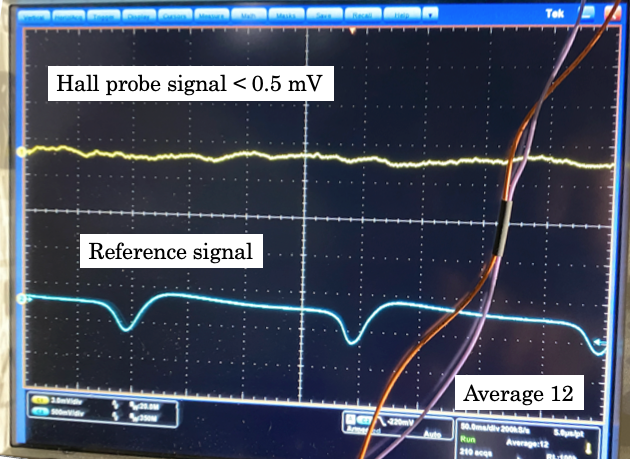}
	\caption{Oscilloscope traces with the compass aligned to the field direction to the level of 0.3 mrad: Hall-probe signal (yellow, amplitude below 0.5~mV) and rotation-phase reference (cyan), averaged over 12 triggers.}
\label{fig:aligned}
\end{figure}

For the field-direction measurement the compass was mounted at the centre of the polarized-target region on a support that allows the orientation of the rotation axis to be adjusted.
The orientation was fine-tuned by minimizing the Hall-probe signal on the oscilloscope, averaged over 12 triggers (Fig.~\ref{fig:aligned}).

The semi-transparent screen was then photographed (Fig.~\ref{fig:optics}).
The compass was then moved aside along the target axis so that the laser beam reached the second screen, 200~cm behind the first, and a second photograph was taken.
Spot coordinates were determined to 1~mm.
These three positions (two on the semi-transparent screen and one on the second screen), together with the mirror-to-screen distance, determine the direction of the rotation axis.

\section{Spinning-field-concentrator compass}

\begin{figure}[ht]
	\centering
	\includegraphics[trim = 0 20 0 20, width=1.0\columnwidth, angle =0 ] {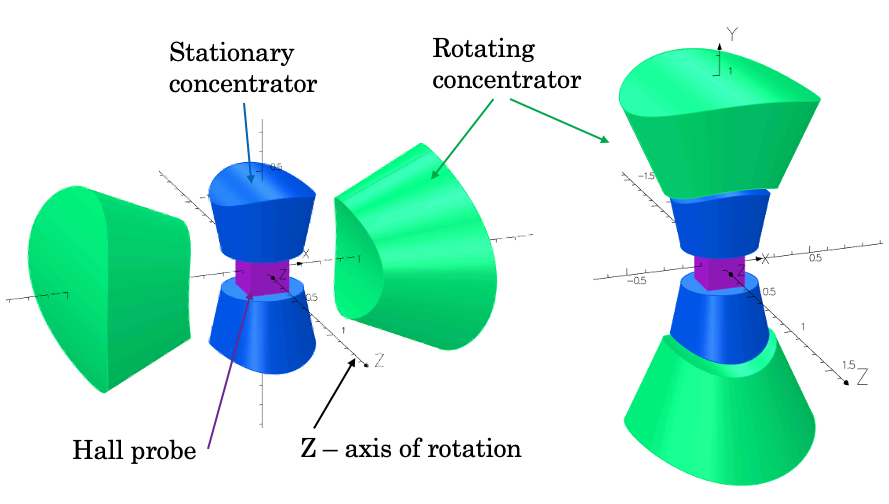}
	\caption{The views of the CAD model of the rotating field concentrator at two rotation phases.}
\label{fig:concentrator}
\end{figure}

The second implementation of the rotation-based compass uses a spinning field concentrator (SFC), Fig.~\ref{fig:concentrator}, see also the patent application~\cite{Patent-2}.
Here the HP (magenta) is stationary and is flanked by a small stationary field concentrator (blue).
A second concentrator (green) rotates about the Z axis, which passes through the HP.
The field enhancement factor (for the modeled geometry) is 4 when the concentrators are aligned (as shown in Fig.~\ref{fig:concentrator}-right) with a drop to 2.5 when the rotatable concentrator is at 90$^\circ$ relative to the stationary concentrator (as shown in Fig.~\ref{fig:concentrator}-left). 
 
The concentrators, made from soft magnetic material e.g. $\mu$-metal, typically produce a residual field at the probe of order 10~mG or less, 
but this is still very large on the scale of the requirement for the compass accuracy. 
In addition, the Hall probe alternating signal from that magnetization can have systematics due to possible difference in the magnetization of two poles.
The practical limit on the symmetry of geometry of the concentrator mounting also leads to a variation of the direction of the magnetic field on the HP and systematic oscillating signal.

\begin{figure}[ht]
	\centering
	\includegraphics[trim = 0 0 0 0, width=1.\columnwidth, angle =0 ] {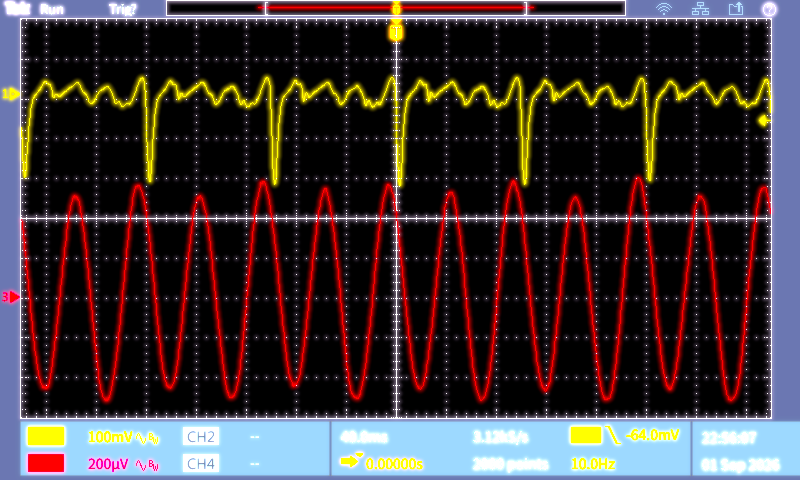}
	\caption{Reference pulses (yellow) and the SFC Hall-probe signal (red) for an external transverse field of 0.25~G. The second harmonic dominates.}
\label{fig:two-harmonics}
\end{figure}

All these limitations are related to the HP signal component which oscillates with the frequency of the concentrator rotation.
At the same time, the HP signal component of interest is due to the external magnetic field (transverse to the axis of rotation).
That part of the HP signal does not change sign with 180$^\circ$ rotation of the rotating concentrator,
but changes its sign with 90$^\circ$ rotation.
For an external transverse field of 0.25~G the HP signal is dominated by the second harmonic (Fig.~\ref{fig:two-harmonics}).

\begin{figure}[ht]
	\centering
	\includegraphics[trim =20 20 20 0, width=1.\columnwidth, angle =0 ] {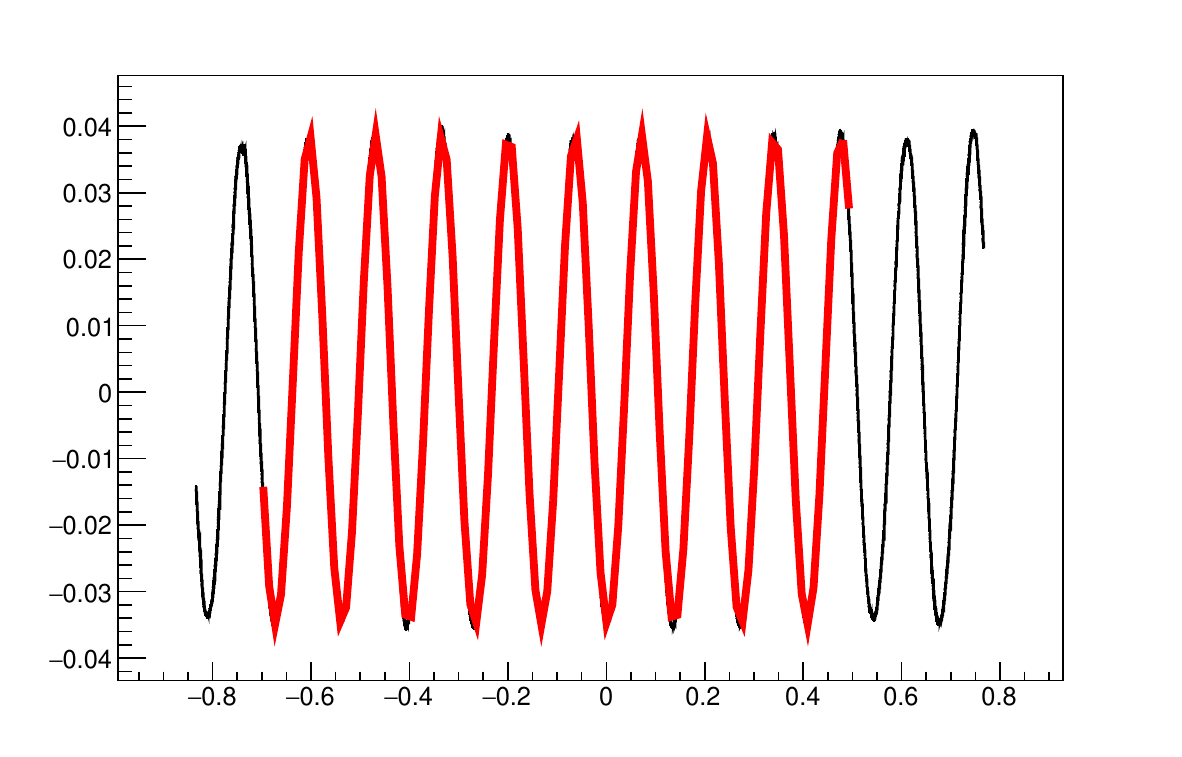}
	\caption{Two-harmonic fit (red) to the SFC signal (black) in a 0.25~G external transverse field. The fit is restricted to the central 9 pulses of the trace (averaged over 32 oscilloscope triggers). The relative uncertainty of the fitted second-harmonic amplitude is $0.7 \cdot 10^{-4}$.}
\label{fig:fit}
\end{figure}

\begin{figure}[ht]
	\centering
	\includegraphics[trim = 0 0 0 0, width=1.\columnwidth, angle =0 ] {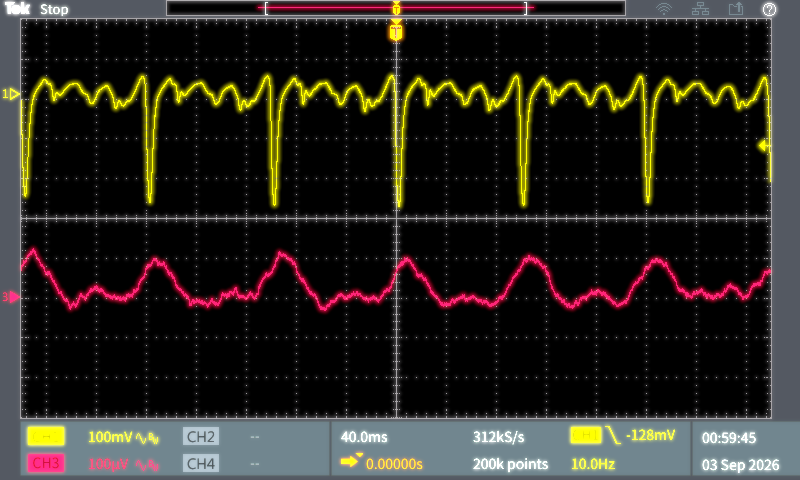}
	\caption{The synchronization pulse (yellow line) and the signal from SFC (red line) for 0.01~G external field. The second harmonic is present.}
\label{fig:low-field}
\end{figure}

A fit of the data with two harmonics, $a\cdot\sin(\Omega \cdot t) + b\cdot\sin(2\Omega \cdot t)$ 
(Fig.~\ref{fig:fit}), gives the second-harmonic amplitude with a relative uncertainty of $0.7\cdot10^{-4}$, corresponding to 0.02~mG.
At a low external field of 0.01~G the second harmonic is still clearly visible (Fig.~\ref{fig:low-field}).

With a single HP this version of the compass is sensitive to only one component of the transverse field, so the alignment is done in one plane at a time.
With two HPs in orthogonal planes the alignment in both planes could be achieved simultaneously.
The SFC compass can operate at much higher rotation frequency than the spinning HP option because it doesn't use the slip-ring contacts and provides higher accuracy for the field direction at the same integration time. 

\begin{figure}[ht]
	\centering
	\includegraphics[trim = 0 20 0 20, width=1.0\columnwidth, angle =0 ] {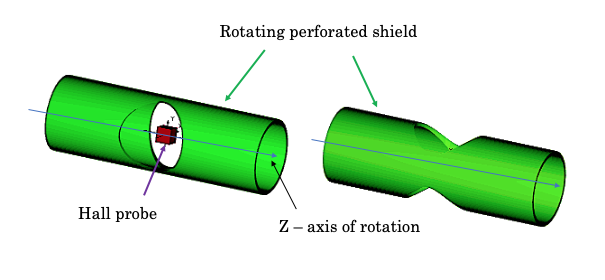}
	\caption{The views of the CAD model of the rotating perforated shield at two rotation phases.}
\label{fig:shield}
\end{figure}

The concept of the SFC could be inverted to a perforated shield, see Fig.~\ref{fig:shield}.
Such a version of the rotating iron redirects the external magnetic field out of the HP with a rotation modulated coefficient.
It is made of a thin wall cylinder ($\mu$-metal) with two large holes in the middle on opposite sides at the location of the HP.
The variation of the magnetic field on the HP with the rotation phase, according to the field calculation with OPERA, is five times less than for the SFC, so the projected accuracy is lower, $\sim$0.35 mrad, but still very good.
The advantage of such a configuration is a much smaller diameter of the whole device, which is essential for applications requiring a compact device.

\section{Gated integrator}
For applications where the use of an oscilloscope is not suitable, the value of the second harmonic of the oscillating signal can be measured by using a gated integrator.
The prototype of such an electronic device has been made.

\begin{figure}[ht]
	\centering
	\includegraphics[trim = 10 10 10 20, width=1.\columnwidth, angle =0 ] {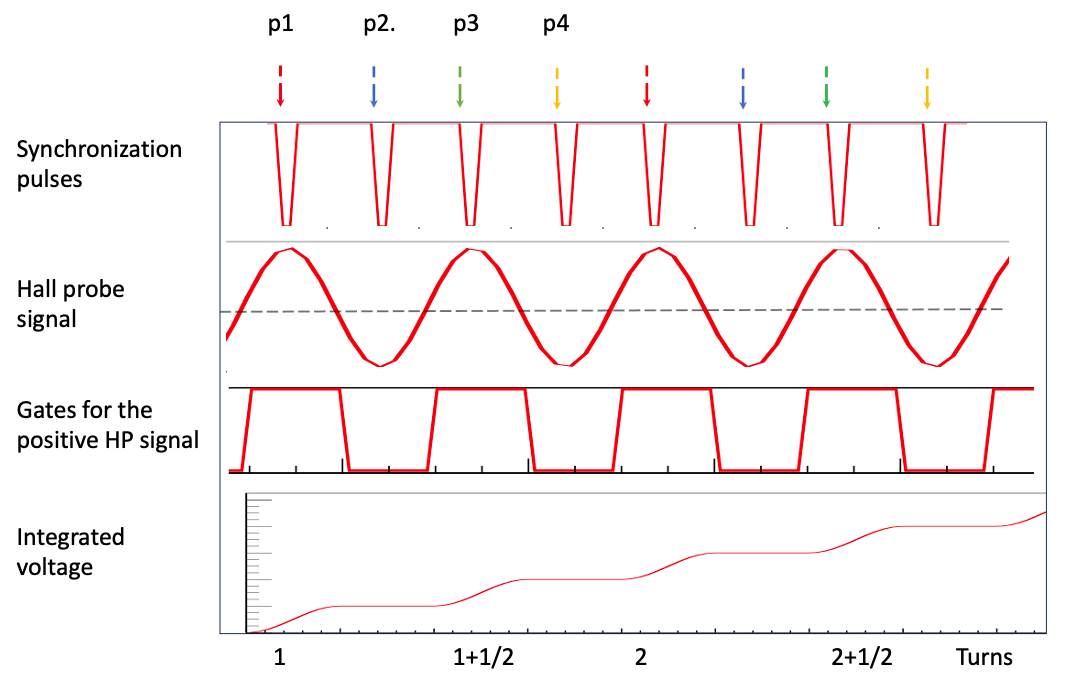}
	\caption{The timing diagram of the gated integrator.
    The top section shows the synchronization pulses, four per turn.
    The second plot from the top shows the alternating signal from the HP.
    The third from the top shows the gates used to select signals with positive polarity of the HP signal.
    The bottom plot shows the integrated voltage from the ``positive'' time intervals at the beginning of the measurement.}
\label{fig:lock-in}
\end{figure}

The timing diagram of the gated integrator for the SFC compass is shown in Fig.~\ref{fig:lock-in}.
Four synchronization pulses are generated per turn, while the Hall-probe signal oscillates at twice the rotation frequency.
The relative phase was optimized by proper orientation of the Hall probe plane.
The gates are generated from the synchronization pulses.
Using gates for both the positive and the negative half-cycles, with the sign inverted for the latter, doubles the detected voltage.
The output used for the compass alignment was integrated with a 10~s time constant, about 100 turns of the SFC.
With a 10~s integration time the output voltage was stable at a level corresponding to a transverse field of 0.02~mG.
This is consistent with the result obtained from the data fit shown in Fig.~\ref{fig:fit}.

\section{Conclusion}

Rotating the Hall probe (or a field concentrator) converts the transverse field component into an alternating signal and allows a high-precision, calibration-free determination of the field direction.
A spinning-Hall-probe compass was built and used in a 25~G field.
The Hall-probe signal minimum was reached at the 0.5~mrad level with only 12 averaged oscilloscope triggers (2~s of data); the overall accuracy of the field direction, about 1~mrad, was set by the optical determination of the rotation axis.
The spinning-field-concentrator compass showed very low noise in the second-harmonic signal, corresponding to a directional resolution of 70~$\mu$rad in a field of 0.25~G.
A custom gated integrator was built to measure the average amplitude of the second harmonic on a digital voltmeter.
This novel compass could find application in many fields, including air and sea navigation, space exploration, and physics experiments.

\section{Acknowledgements}	

We are grateful to G.~Cates, M.~Jones, and R.~Wines for their interest and support of the project.
We appreciate the contributions by C.~Gould, D.~Spiers, A.~Stepanyan, and A.~Tadepalli in the compass development and measurements.
This material is based upon work supported by the U.S. Department of Energy, Office of Science, Office of Nuclear Physics under Contract No. 89243126CSC000213 (TJNAF) and DE-FG02-03ER41240 (N.L.).

\nocite{*}
\bibliographystyle{apsrev4-1}
\bibliography{aipmain.bib}
	
\end{document}